# A time projection chamber with optical readout for charged particle track structure imaging

U.Titt[*+], A.Breskin[#], R.Chechik[#], V.Dangendorf[*], H.Schmidt-Böcking[+], H.Schuhmacher[*]

[*] *Physikalisch-Technische Bundesanstalt, 38116 Braunschweig, Germany*
[+] *Universität Frankfurt, 60486 Frankfurt, Germany*
[#] *Weizmann Institute of Science, 76100 Rehovot, Israel*

**Abstract**
We report about a nuclear track imaging system which is designed to study in detail the ionization topology of charged particle tracks in a low-pressure gas. The detection method is based on a time projection chamber (TPC) filled with low-pressure triethylamine (TEA). Ionization electrons produced by energetic charged particles are three-dimensionally imaged by recording light from electron avalanches with an intensified CCD system. The detector permits to investigate the spatial ionization distributions of particle tracks in gas, of equivalent length and resolution in tissue of 4 µm and 40 nm (RMS), respectively. We explain the relevance of this technique for dosimetry, describe the experimental method and the basic operation parameters. First results of the chamber response to protons and alpha particles are presented.

**Corresponding author:**
Volker Dangendorf                      email: volker.dangendorf @ ptb.de
Physikalisch-Technische Bundesanstalt   Tel:   ++49 (0) 531 592 7525
Bundesallee 100                         Fax:   ++49 (0) 531 592 7015
38176 Braunschweig

# 1. Introduction

Microscopic information about the ionization density along tracks of charged particles is required in various areas. Examples are microdosimetry, e.g. for understanding radiation-induced effects in organisms and semiconductor material, and radiation protection dosimetry. A generally accepted model which can predict radiobiological effects from the physics of radiation interaction does not exist to date. The models under discussion are based on microscopic information about radiation interaction with cells and relevant structures therein and therefore require interaction data down to the nanometer level (examples are restricted linear energy transfer, energy imparted, proximity functions). These data are commonly provided by track structure calculations which have to rely on very scarce and poorly known atomic and molecular cross sections.

In general, experimental methods for particle track analysis are based on the simulation of a microscopic volume in a solid or liquid by replacing it by a much larger cavity filled with gas of much lower density. Although the interaction data determined in gases might not directly be relevant for radiation effects induced in liquids or solids, they represent important benchmarks for track structure calculations. Hence, any experimental method able to measure interaction data down to the nanometer scale contributes to the understanding of radiation action in matter.

In radiation protection dosimetry the present concept is based on "protection quantities", used to estimate the risk of an irradiated person, and "operational quantities", used for measurements in health physics [1, 2]. The definition of operational quantities, which serve as conservative estimates for the protection quantities in the daily routine, implies specific irradiation conditions with respect to the radiation field and to the receptor, i.e. the body or a so-called phantom. It takes into account the biological effectiveness of the different types of radiation by quality factors which are defined as a function of the linear energy transfer, LET, of charged particles at a position inside the phantom. In contrast, the protection quantities are to be determined for the irradiated person and the biological effectiveness of different types of radiation and the radiation sensitivity of different organs is taken into account by radiation weighting factors (defined as a function of the type and energy of the radiation outside the body) and tissue weighting factors, respectively.

No methods exist to date for determining these dosimetric quantities directly according to their definition. Instead, different and, at least partly, complicated methods are employed. Examples are radiation field analysis by spectrometric methods or the use of a low-pressure tissue-equivalent proportional counter (TEPC). The latter is able to simulate a tissue volume of about 1 µm in diameter and to measure a quantity which is closely related to the LET as long as the charged particles have ranges of at least several micrometers in tissue. However, for short-ranged charged particles, e.g. those produced by low-energy neutrons, dose equivalent is significantly underestimated [3].

The particle track chamber may address these problems in various ways. On the one hand, the particle track chamber allows to identify the type and energy of charged particles produced in its wall and gas and therefore enables the analysis of the primary radiation field. It also allows to measure the neutron energy distribution directly, using a $^3$He-gas filling. On the other hand, the chamber allows a more accurate measurement of LET than a TEPC because the above-mentioned restrictions of a TEPC do not apply to the track chamber.

In recent years, several attempts to apply charged particle track imaging in dosimetry, microdosimetry and neutron spectroscopy were reported. Turner et al. [4 - 6] employed a method, where the ionization electrons along a particle track are accelerated by a damped high voltage radio frequency and cause scintillation light. The scintillation is viewed by two cameras from orthogonal sides and assembled to a three-dimensional picture. The advantage of this method is that the ionization electrons produce their light immediately at the place of their generation, thus avoiding possible inaccuracies due to electron diffusion. However, due to the three-dimensional extension of the charge distribution in the sensitive volume, the optical system requires a large depth of field (possibly several centimeters), and compromises between resolution and light sensitivity cannot be avoided.

Another approach was reported by Marshall et al. [7,8]. Their method is based on the well-known cloud chamber technique to visualize particle tracks in water vapour. An excellent spatial resolution of less than 5 nm (scaled to materials of a density of 1g/cm$^3$) was obtained, making this method the first choice in fundamental microdosimetric measurements, where high rate and real time capability are not required. The drawback of this method is the extremely long duty cycle due to the long recovery time of the chamber.

More recent approaches have been reported by Breskin et al, electronically recording the trails of individual charges deposited by particles in low-pressure gases. These methods comprise single electron counters [9-11], and single ion detection techniques [12]. The latter method should permit to measure equivalent resolutions in tissue of the order of less than one nanometer.

In the present work we will describe the concept, the instrumental setup and first experimental results of a newly designed OPtical Avalanche Chamber (OPAC) which will be applied as particle track imaging device for dosimetric applications[13]. It is based on a time projection chamber with a three-dimensional optical imaging capability of particle-induced ionization patterns, by recording the light emitted from electron avalanches with an intensified CCD camera.

We report on the evaluation of the basic operation parameters such as:
- light and charge production in low-pressure triethylamine (TEA) in parallel gap gas amplification structures,
- single electron detection efficiencies under various amplification conditions,
- drift and diffusion of electrons in low-pressure TEA.

Finally, preliminary results of the chamber response to well-defined particle irradiation are presented.

**2. Concept**
Previous works in the field of optically read out avalanche counters were reported in the 1980s by CERN and Weizmann Institute groups (see e.g. [14,15] and related references therein). As opposed to the conventional electronic detector readout, the optical method focuses on the capture of the UV or visible light emission, which in some gases copiously accompanies the formation of charge avalanches. The successful implementation of this readout scheme was facilitated by the availability of gateable, image-intensified CCD cameras (IICCD) at moderate costs.

Compared to electronic readout, the optical method has some advantages:
- It offers low cost, high resolution, genuine two-dimensional multi-hit readout capability.
- It is insensitive to electronic noise or RF-pickup signals in the detector environment.
- With properly chosen gases mesh electrodes can be applied instead of wires, which makes the chamber simpler and more rugged.

The development of suitable gas mixtures emitting light in the near UV range and having a rather large light yield per avalanche electron made it possible to operate those chambers not only in the saturated pulse mode [16] but also in a steady-field, charge-proportional mode. Only in the latter operation mode does the registered light pattern contain the information about the ionization energy locally deposited into the chamber gas. Moreover, the steady-field operation enables the optical time projection technique [17] to be used because the time development of the light pulse is always proportional to the amount of electrons arriving at the amplification structure in a certain time interval. In the optical time projection chamber the light emission always occurs in the same layer. Therefore no depth of field of the optical system is required, and the maximum available optical aperture can be used. In addition, one obtains spatial information about the charge distribution perpendicular to the amplification plane (see fig. 1. and chapter 3 for details of the instrumental setup).

A key element in the optical readout technique is a gas mixture with sufficient light yield in the visible or near UV range to make the avalanche detectable with the IICCD. It was shown [18-21] that gas mixtures containing tetrakis[dimethyl-amino]ethylene (TMAE) and triethylamine (TEA) vapours enable operation of parallel grid chambers with large light yield in the proportional mode. TMAE-based gas mixtures have the advantage of emitting visible light at a peak wavelength of $l = 480$ nm [22]. However, TMAE handling is rather delicate since its vapour pressure is small and TMAE partial pressure above several hPa usually requires heating of the detector and its gas supply facilities [23]. Furthermore, due to photon feedback, the diameter of a single electron avalanche in low-pressure TMAE mixtures extends to a few mm [15]. TEA, on the other hand, emits in the near UV (peak emission at $\lambda = 280$ nm [19, 22]) which can still be comfortably handled with regular quartz and Al-mirror optics. The light yield of electron avalanches in TEA mixtures is about 10 times larger than with TMAE and the single electron spot size is below 1 mm even at 10 hPa pressure (see below). Its vapour pressure at normal temperature is sufficiently high (approximately 120 hPa at 300 K [20]) to operate the detector and the gas supply system at room temperature. For low pressure operation (up to 60 hPa), commercial techniques can be used to control the TEA flow and pressure. The IICCD readout provides the two-dimensional images of the electron distribution. In fact, only the projection of the ionization track onto the plane between the amplification grids is recorded. As was proposed by Charpak et al [14], and later systematically developed by Breskin et al [9-11], an additional spatial information can be obtained by recording the temporal flow of charge into the amplification region. The known drift velocity in the interaction volume of the chamber and the measurement of the arrival times of electron clusters relative to an event trigger provides the position of the cluster origin in the coordinate axis perpendicular to the CCD image plane. In the proportional amplification area, this charge flow is transformed into a time-dependent light emission with an intensity proportional to the charge. Thus, three-dimensional localization of each electron or electron cluster is achieved by combining the CCD-image and the time-resolved information of the light emission e.g. with photomultiplier tubes [17].

## 3. Experimental Setup
### 3.1 The detector
The OPAC consists of a parallel-field drift chamber of 10 cm in diameter and length. It is mounted in a stainless steel vessel with a quartz window light-output port close to a parallel plate light amplification region. The housing provides three beam ports to allow controlled irradiation of the sensitive volume with charged particles at angles of 90°, 45° and 0° with respect to the image projection plane. The volume is filled with the vapour of $(H_5C_2)_3N$ (TEA) at a pressure of 10 hPa. Charged particles crossing this volume produce free electrons along their path, which drift towards the amplification stage under a homogeneous electric field (see fig. 1). The drift cathode (DC) and the drift anode (DA) consist of stainless steel wire meshes of 81% optical transmission. The parallel drift field is shaped by a series of guard electrodes made of stainless steel rings.

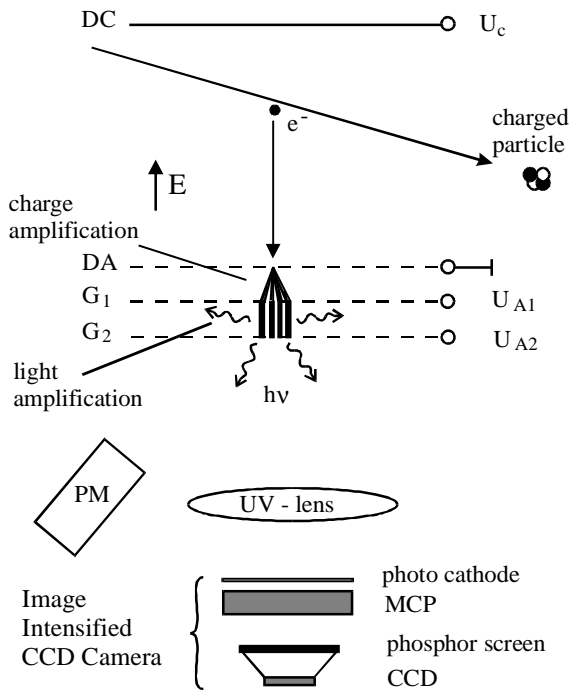

**Fig.1** *Schematic view of the OPtical Avalanche Chamber (OPAC) with drift cathode (DC), drift anode (DA) and amplification grids $G_1$ and $G_2$, and its optical readout elements UV lens, photocathode, multi-channel plate (MCP), charge-coupled device (CCD) and photomultiplier (PM). In the first setup only single step amplification between DA and $G_1$ was used. To achieve higher light output a second amplification stage between $G_1$ and $G_2$ was implemented.*

From the drift region, which defines the sensitive volume of the chamber, the charges enter an amplification zone. The detector has a two-step homogenous electric field amplification system, consisting of a stack of grids arranged at a spacing of 3.2 mm. Electrons which drift into the amplification region are multiplied in a first proportional charge amplification stage between DA and $G_1$. This stage is followed by a light amplification region, where a rather low reduced electric field of $E/p \approx 65$ Vcm$^{-1}$hPa$^{-1}$, applied between $G_1$ and $G_2$, allows proportional scintillation induced by the preamplified electron swarm [23].

### 3.2 The Optical readout system
With the technique of a parallel drift field and double-step charge and light amplification structure, a two-dimensional image of the track is projected onto the mesh of the light amplification stage. This image is recorded by a UV-sensitive optical imaging system, consisting of three main parts:

A UV lens focuses the scintillation light pattern onto the photocathode of an image-intensified CCD camera (IICCD). The lens is an Al mirror lens (NYE, Lyman alpha) with a focal length $f = 90$ mm and an aperture of $F = 1.1$. Despite the disadvantage of a pillow-like distortion of the image by this lens it was selected due to its large $F$ value. With an object of 90 mm in diameter, corresponding to the sensitive part of the chamber, and an image size of 25 mm, an effective solid angle of $\Delta\Omega/4\pi = 2.44 \cdot 10^{-3}$ is calculated.

The conversion of UV light into photoelectrons in the image intensifier (Proxifier BV 2562 B$_c$Z, [24]) is made by a bialkali photocathode, 25 mm in diameter. A bialkali photocathode was chosen due to its lower thermal noise as compared to the usual multialkali photocathodes.

The photocathode can be pulsed with short (down to 10 ns) high-voltage pulses. This allows controlled and very short exposure times of the CCD and avoids accumulation of noise as well as multiple exposures of the CCD. The electrons from the photocathode are multiplied by a single-step multi-channel plate (MCP, BV2562QG) and accelerated in a parallel electric field. The accelerated electrons hit a P-43 phosphor screen which emits visible light with a decay time of 1 ms. The maximum light amplification of the image intensifier is $2 \cdot 10^5$. The phosphor is coupled by a conical fiber-optical faceplate to the CCD of a standard slow-scan camera (EG&G, M4005).

The spatial resolution of the optical system is quoted to be 38 line pairs per mm, which transforms to a resolution of 10.5 line pairs per mm in the object plane. This resolution does not limit the resolution of the instrument as larger effects are contributed by the electron transport in the gas (see chapter 4.2).

To avoid the accumulation of dark noise between two events, the CCD is continuously cleared after the readout of an event until the trigger of a new event occurs. With this technique the noise induced by the CCD dark current is negligible. When receiving an event trigger, the photocathode is pulsed with a high-voltage pulse ( 200 V ) of about 3 µs duration, corresponding to the maximum possible drift duration of the electrons from a particle track.

The secondary light emission as a function of time contains the spatial information of the ionization density distribution perpendicular to the drift projection plane. This light transient is observed by six photomultipliers (PHILIPS XP2018QB) whose anode outputs are connected in three groups. By analysing the time dependent signal ratios of the photomultipliers a rough information on the position in the projection plane can be derived. With this method the ambiguity in the sign of the angle between the particle track and the image projection plane is removed.

### 3.3. The data acquisition and analysis system

The CCD images are transferred to a data acquisition system via an 8 bit variable scan frame-grabber card. The maximum frame rate of the present system is restricted by the camera to 15 s$^{-1}$. The anode signals of the photomultipliers are recorded with a digitizing oscilloscope and the data is transferred to a computer.

The data acquisition and analysis system has to perform the following tasks:
1. collection of the digitized CCD and photomultiplier data,
2. optional storage and display of both data sets,
3. reconstruction of the three-dimensional ionization distributions,
4. track detection and histogramming of longitudinal and lateral ionization density profiles,
5. identification of particle type and energy, and generation of frequency distributions,
6. extraction of absorbed dose and track-structure data.

In the current version of the acquisition system, items 1, 2 and 4 have been implemented. Thus we are able to collect and store the complete three-dimensional information obtained by the sensor. Presently, the response of the OPAC to various particles at different energies is studied to develop the data reduction algorithms for items 3, 5 and 6.

### 3.4. The gas supply system

The OPAC has been operated at 10 hPa TEA in a proportional scintillation mode. To prevent deterioration of the gas quality due to leakes, outgassing of detector materials and avalanche-induced gas ageing the detector was continuously supplied with fresh gas. A dedicated gas supply system was developed for this purpose and is schematically shown in fig. 2. It consists of a glass reservoir for the liquid TEA with the possibility of bubbling carrier gases through the liquid.

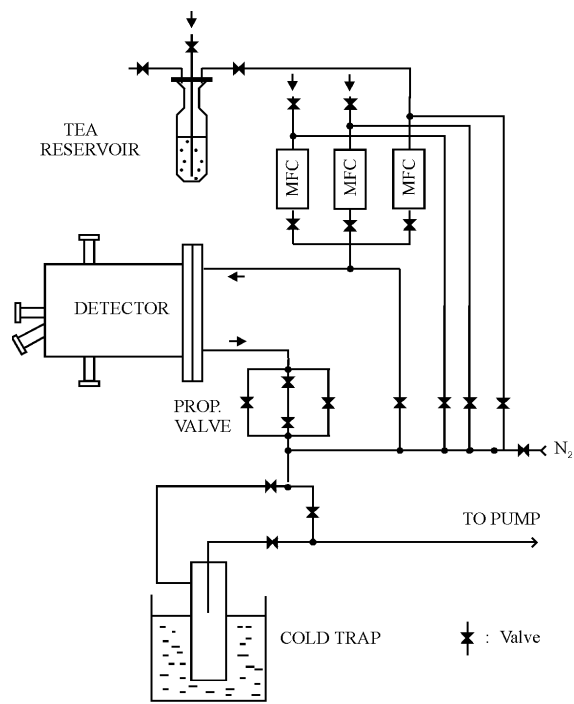

**Fig.2** *Schematic drawing of the gas supply system. It is equipped with 3 gas flow controllers (MFC) for producing gas mixtures of up to 4 components for approximating the atomic composition of tissue (TE gas).*

In this operating mode, the partial pressure of TEA can be controlled by adjusting the temperature of the TEA bubbler. However, for low-pressure operation (below 50 hPa) of the detector, the TEA partial pressure can also be controlled by standard mass flow controllers (MFC, MKS 1259CC). For mixtures of TEA with other gases, two additional MFCs are available. This option will be required in the future for using tissue-equivalent (TE) gases. Pressure control is realized by measuring the chamber pressure with a capacitive pressure gauge (MKS 622) and a proportional valve (MKS 248) mounted behind the detector, downstream to the pumping line. To avoid poisoning of the pump oil and the environment with TEA vapours and to prevent backdiffusion of oil vapours into the detector a liquid nitrogen trap is mounted between the control valve and the roughing pump.

The operating gas pressure (*p*) of 10 hPa was chosen in order to achieve good position resolution as well as stable operation conditions at high charge and light gain.

The position resolution at tissue density ($s_h$) is determined by the scaling factor of the gas density ($r_g$) to tissue density ($r_h = 1$ g·cm$^{-3}$) and the position resolution of the detector ($s_g$):

$$s_h = s_g \cdot \frac{r_g}{r_h}$$

Due to the proportionality of $r_g$ and gas pressure *p*, one should aim for the lowest possible pressure in order to get maximum resolution. However, the resolution obtainable in the gas is limited by charge diffusion during the drift process and varies like $s_g \propto p^{-1/2}$. Therefore, the resolution $s_h$ obtainable in tissue varies as $s_h \propto p^{1/2}$, and one should aim at the lowest possible pressure in order to get the best resolution.

High gain operation of low pressure isobutane- and methane-filled detectors was investigated by Breskin et al [25, 26]. It was shown that at pressures below about 10 -15 hPa the maximum attainable gain drops due to feedback effects like photon- and ion-induced secondary avalanches. We investigated the maximum attainable gain for TEA in a pressure range between 5 and 20 hPa. For single primary electrons stable operating conditions at a gain of up to 10$^6$ in a single parallel gap amplifications stage was achieved at a pressure around 10 hPa. Hence, this pressure was chosen as a reasonable compromise between the desired resolution and sufficient amplification. TEA has a vapour pressure of about 120 hPa at 298 K which is sufficient for operating the chamber at 10 hPa with the mass flow controller and without the necessity of heating the gas supply components to prevent condensation. The scaling factor to 1 g·cm$^{-3}$, which can be achieved at 10 hPa TEA ($r_g = 41.7$ μg/cm$^3$), is about 2.5·10$^4$. Hence a resolution of 1 mm in the detector scales down to about 40 nm in tissue.

## 4. Results and Discussion
### 4.1. Charge and light amplification
In the past, only mixtures of noble gases and hydrocarbons with small concentrations of TEA were investigated and applied [14,15]. For our work the total light and charge yield in pure low-pressure TEA had therefore to be investigated. In order to compare the performance of pure low-pressure TEA with that of other gas mixtures and to find suitable operation parameters, the ratio of light to charge (photons per avalanche electron) was measured. Details on the measurement procedure can be found in [13].

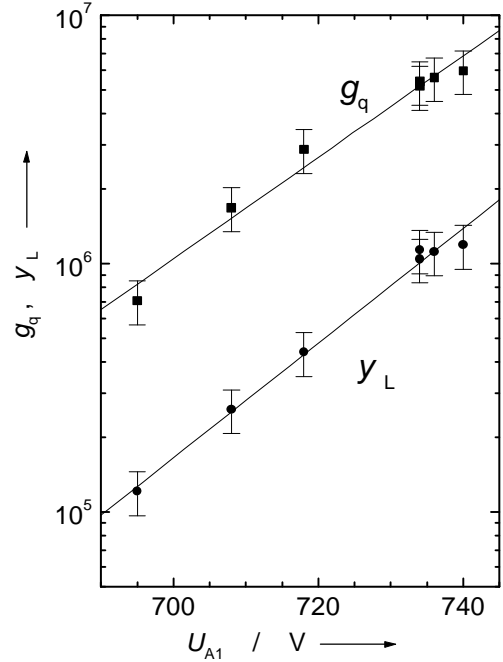

**Fig.3** *Charge amplification $g_q$ and photon yield $y_L$ in a single 3.2 mm parallel amplification gap as a function of the applied voltage $U_{A1}$, measured at 12 hPa TEA. The avalanche was induced by single photoelectrons*

Fig. 3 shows the total photon yield per initial electron ($y_L$), and the charge gain ($g_q$), as a function of the applied voltage in a single parallel-plate amplification gap of 3.2 mm width in 12 hPa TEA. Fig. 4 presents the light-per-charge yield as function of the TEA pressure at two fixed reduced electric fields. It should be noticed that the light-per-charge ratio is higher for lower reduced electric fields. This can be explained by the fact that, with increasing field the average collision energy of the electrons with TEA molecules increases. With increasing collision energy the ionization probability of the TEA molecules increases more rapidly than the excitation probability [23]. Excitation of TEA molecules by electron collisions is the main cause for the scintillation process.

Along the central part of a non relativistic heavy ion track the ionization density is generally sufficiently high so that even at a moderate light gain of the avalanche chamber the optical system is able to detect the track. However, tracks of minimum ionizing particles or of fast delta electrons at some distance from the core of a heavy ion track mainly consist of separated single electrons. A comparison of the experimental data with Monte Carlo simulations [13] showed that with a single amplification step the obtainable photon yield would not be sufficient for efficient detection of these single electrons (see also chapter 4.5). Therefore a second amplification stage was implemented, which, depending on the ionization density of the particles investigated, was operated at a reduced electric field between 22 and 81 $Vcm^{-1}hPa^{-1}$. Due to this element the light output could be augmented by a factor of up to 20. Fig. 5 shows the light amplification of the second 3.2 mm parallel gap as a function of the applied voltage, measured in 10 hPa TEA.

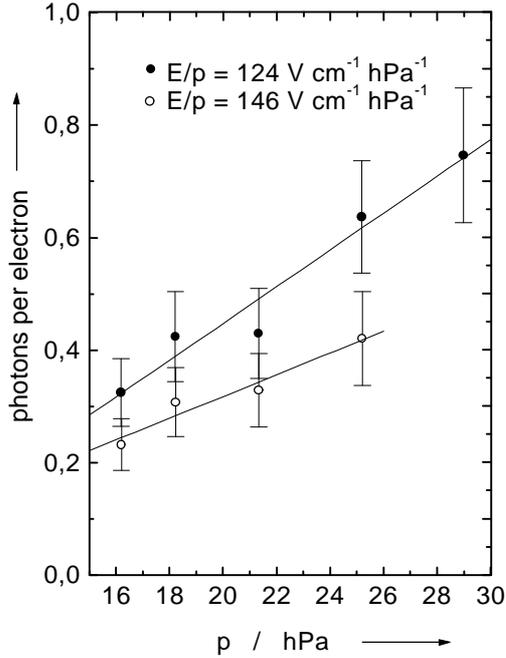

**Fig.4** *Dependence of light per charge ratio on the TEA pressure in a single-step parallel gap of 3.2 mm in width at two fixed reduced electric fields*

**4.2. Electron drift in low-pressure TEA**

The electron transport parameters in the drift cell determine the obtainable localization resolution of the detector. The transverse diffusion mainly causes the uncertainty in the position measurement of the projected image in the CCD object plane. Along the drift direction the localization measurement is mainly influenced by the transition time of electrons through the second light amplification gap and by the longitudinal diffusion. Since no data are available for the electron transport parameters in TEA we measured the drift velocity, the transverse and longitudinal diffusion and the time width of the light burst induced by a point source of slow electrons in the single- and double-step amplification modes.

To measure the drift parameters (drift velocity and lateral diffusion) of single electrons, free electrons were ejected from a small spot on the drift cathode. The electrons were produced by a 3 ns long pulsed UV laser beam (nitrogen laser, emission wavelength $\lambda = 337$ nm) hitting a small aperture of 0.1 mm in diameter in an aluminium drift cathode. The UV-light emission of the laser was registered in the PM and used as time reference in all measurements. The photoelectrons produced by the laser in the Al wall of the aperture were transferred by an electric guidance field into the drift zone. To assure single electron injection, the laser intensity and the guidance field were adjusted such that only one out of ten UV pulses resulted in a free electron, visible as a pulse in the detector readout electronics.

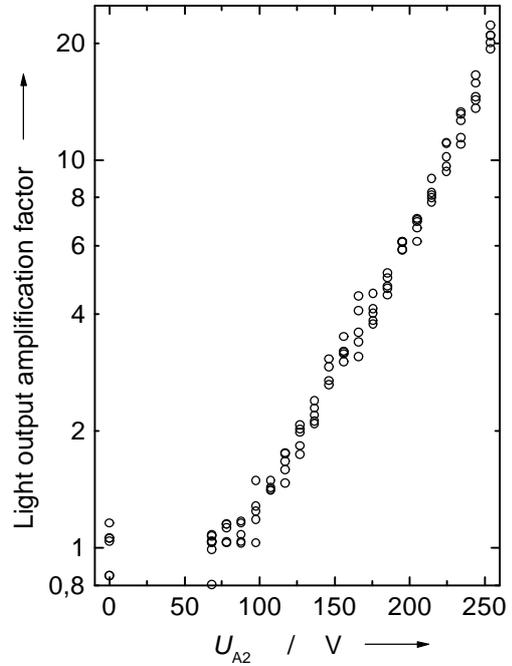

**Fig.5** *Dependence of light yield on the applied-voltage $U_{A2}$ in the second parallel field-proportional scintillation stage 3.2 mm in width in TEA at 10 hPa pressure*

**a) Drift velocity**

The drift time as a function of the drift field strength was determined by measuring the delay between the UV laser pulse in the PM and the secondary light emission induced by the avalanche electrons in the amplification stage, for variable reduced electric drift field strength values. The pressure was kept at 10 hPa in all measurements. The resulting electron drift velocity is shown in fig. 6. In the relevant *E/p*-range, the drift velocity varies almost linearly with *E/p*, with two different slopes. Even at the highest field no saturation of the drift velocity is observed. The maximum applied field was limited here by spurious discharges within the detector.

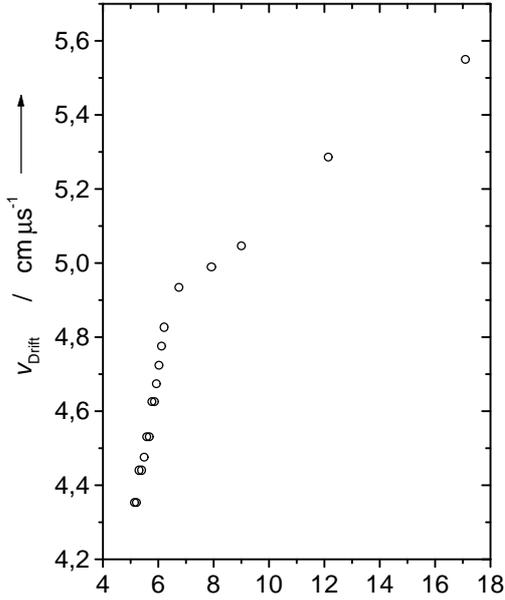

**Fig.6** *Drift velocity of electrons in 10 hPa TEA as a function of the reduced drift field strength.*

**b) Transverse diffusion ($s_T$)**

To measure lateral diffusion, the images of single electron induced scintillation in the amplification region were recorded with the IICCD. For different drift field values the centre of gravity of each single electron spot in a series of several thousand events were calculated. The standard deviation ($s_T$) of the transverse diffusion for 10 cm drift in 10 hPa TEA is shown in fig. 7 as a function of the reduced drift field strength. The transverse diffusion reaches a flat minimum at a reduced electric field of about 5 Vcm$^{-1}$hPa$^{-1}$. This corresponds to a resolution of about 120 nm in tissue for an electron drift path length of 10 cm. For dosimetric applications and particle identification, this is a sufficiently good value. In microdosimetry, however, only resolutions of a few tens of nm would constitute a significant improvement compared to existing technologies like TEPCs. A reduction of the drift length to 1 cm could improve the obtainable resolution in gas to 1 mm, corresponding to about 40 nm in tissue density.

The measurement of the lateral width of a particle track is determined by the vector component of the transversal diffusion, which points into the direction perpendicular to the particles' flight direction and perpendicular to the drift field lines. This component of the transversal diffusion is plotted as $s_P$ in fig. 7.

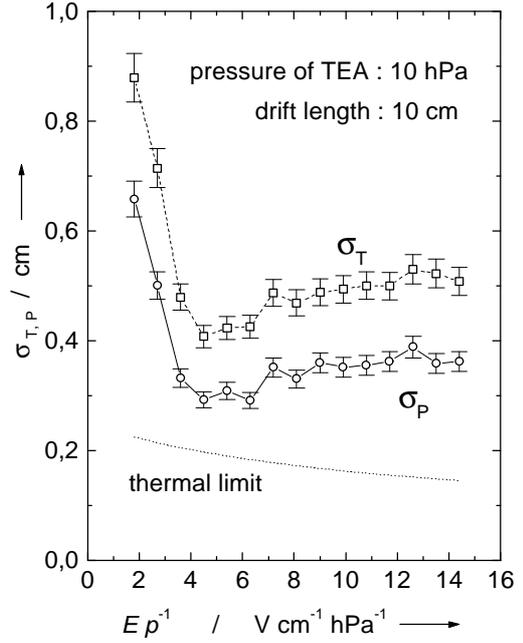

**Fig.7** *Transverse ($s_T$) and projected ($s_P$) diffusion of electrons in 10 hPa TEA for 10 cm drift length as function of the reduced electric field strength.*

**c) Longitudinal diffusion and time spread of light emission**

The spatial resolution in the direction parallel to the drift field lines depends on the longitudinal diffusion and on the time spread of the light emission. To determine the two contributions separately, the width of the light and charge pulses from the UV laser induced avalanches were measured. The laser was positioned in front of the quartz window in such a way, that an intense laser beam hits the amplification grids and the drift cathode and causes simultaneously the emission of an electron cluster from each electrode. Hence, the charge signal, collected at $G_2$, provides two temporal components: an instantaneous one caused by the photoelectrons from the first amplification grids, and a delayed one, caused by the photoelectrons from the drift cathode, 10 cm away from the amplification structure. The first pulse contains only the diffusion and other fluctuation effects of the avalanche formation while the time spread of the delayed pulse is additionally influenced by the longitudinal diffusion along the homogeneous drift space. Thus, the longitudinal diffusion in time is derived from these measurements by quadratic subtraction of the width of the two components of the charge signal trail following each laser shot. The contribution of the width of the laser pulse (3ns) can be neglected. The geometrical longitudinal diffusion is calculated using the measured drift velocities (see section 3a). Fig. 8 shows the longitudinal diffusion ($s_L$) for various drift field values.

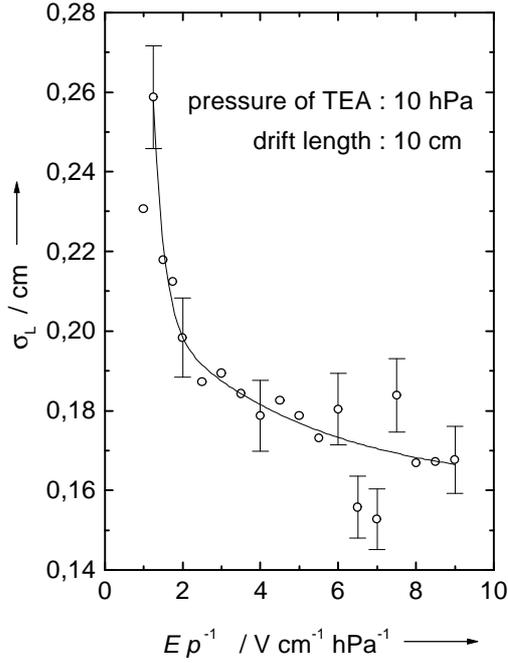

**Fig.8** *Longitudinal diffusion of electrons in 10 hPa TEA for 10 cm drift length as function of the reduced electric field strength.*

The combination of longitudinal diffusion and the time structure of the avalanche formation determines the spatial resolution in the drift direction. For a 10 cm long drift path at a typically applied reduced drift field of about 5.5 V cm$^{-1}$ hPa$^{-1}$, an RMS time spread of 60 ns [± 10 %] was measured. With a drift velocity of 4.4 cm µs$^{-1}$, this leads to an overall spatial uncertainty of 2.7 mm in the drift direction, which corresponds to 112 nm at tissue density.

### 4.3. Single electron spot size

An advantage of TEA as a scintillating vapour is its relatively large absorption efficiency for avalanche induced photons. These photons are isotropically emitted from the avalanche, giving rise to secondary avalanches surrounding the primary avalanche core and blurring the primary avalanche image. This is the main cause for the relatively large lateral size of a single electron avalanche spot in the CCD image when using TMAE at room temperature [15]. We measured the diameter of single electron spots in low-pressure TEA operated detectors with a single step amplification system which resulted in a width of about 0.4 mm (fwhm).

### 4.4. Single electron detection efficiency

The detection efficiency for single electrons was measured with a setup similar to that used for measuring the drift velocity and the transverse dif-fusion, i.e. with single electrons photo-ionized by a UV-laser pulse from an Al photocathode. The photocathode of the IICCD was gated by the signal of the laser pulse, recorded with the photomultipliers.

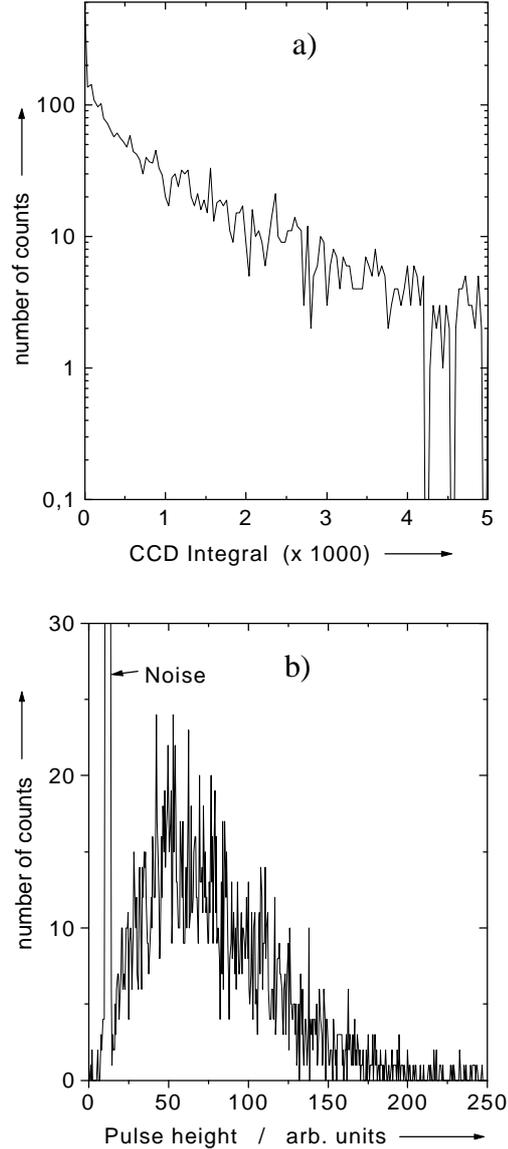

**Fig.9** *Light (a) and charge (b) pulse height distributions of avalanches from single primary electrons in 10 hPa TEA at $U_{A1}$ = 530 V in the first gap and $U_{A2}$ = 844 V in the second gap.*

The reference for the occurrence of a single electron emission from the photocathode was provided by the electronic signal induced on the amplification anode. Fig. 9 shows light (9a) and charge (9b) pulse height distributions. The charge pulses are clearly separated from the noise, thus a detection efficiency of close to 100 % in the charge channel can be assumed. Therefore we can compare the number of detected secondary light pulses through the IICCD with the number of charge pulses above the noise threshold.

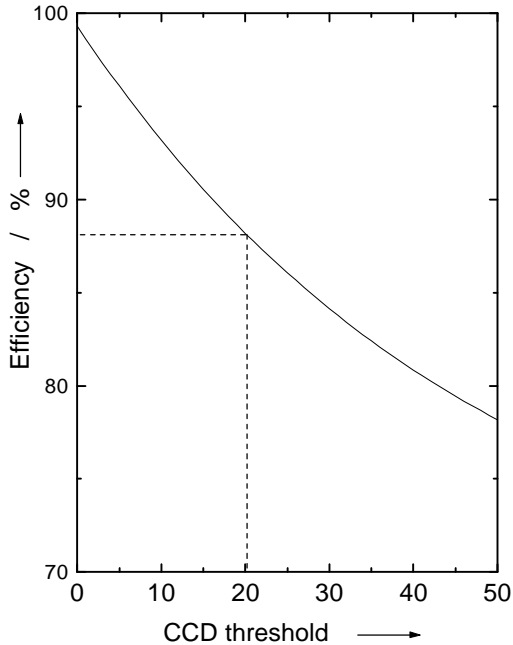

**Fig.10** *Single electron detection efficiency of the optical system (p=10 hPa, $U_{A1}$=530 V, $U_{A2}$=844 V). The CCD threshold relates to the sum of all pixel contents. If a threshold value of 20 (compared to the maximum single pixel value of*

The ratio of the number of events detected optically to the number detected electronically directly gives the single-electron detection efficiency of the OPAC at given amplification parameters. Fig. 10 shows the measured results as a function of the CCD threshold. The pixel sum (sum of the content of all pixels) of the image intensifier and CCD for events induced by noise at normal temperature is below 20 counts (at a lower threshold for the CCD of 13 mV at 2 V full range ). Therefore, above a threshold of the CCD pixel sum of 20 counts the images do not contain noise while the single electron detection efficiency approaches 87 % .

**4.5 Track imaging results**
Images of α-particle tracks from a $^{241}$Am source were recorded with the OPAC. α-particles crossed the drift volume perpendicular to the drift field lines at a distance of 5 cm to the charge amplification gap. A Si(Li) detector of 200 mm$^2$ sensitive area, mounted on the opposite side of the drift cell, was used as an active collimator and as a trigger for data acquisition. Fig. 11 shows examples of α-particle tracks obtained with the OPAC in a single- (11a) and in a double-step (11b) amplification modes. 5 MeV α-particles produce about 800 electrons per centimeter path length in 10 hPa TEA.

Most of these electrons are located around the central core of the track of less than 1 millimeter lateral width. Therefore, in this central part of the track the recorded light pattern is composed of many overlapping single electron avalanches which cannot be individually resolved by our system. Only at some distance from the track core, where the ionization pattern is determined by few sparsely ionizing δ-electrons, single electrons can be resolved.

Fig. 12a shows the lateral projection of the ionization distribution of the image of fig. 11a. With one amplification step the detection efficiency for avalanches originating from single electrons is very low, as was explained in section 4.1. Only double-step amplification provides close to 100 % single electron detection efficiency. For comparison fig. 12b and 12c show the lateral and longitudinal projections of the event in fig. 11b with double step amplification. In 12b the "shoulders" produced by fast delta electrons can clearly be observed**.** The decreasing sensitivity towards the edge of the image, which is seen in the figures 11 and 12c is due to the distortion caused by the UV lens (see section 3.2). Nevertheless, from the undistorted central part of the image it is obvious that even for densely ionizing ions like α-particles the stochastic nature of energy deposition along the central particle track is still visible in a low-density medium.

We are presently investigating the response of the OPAC to particles of known type and energy. Data for protons, α-particles, nitrogen, argon and krypton ions with energies from 500 keV/u to 29 MeV/u were obtained at different incidence angles. Fig. 13 shows some examples of particle tracks with particularly interesting ionisation patterns due to fast secondary particles (δ-electrons, recoil ions). These particles are ejected by the primary ion and produce themselves extended ionization tracks. For all measurements we recorded and stored the complete three-dimensional track-structure information for each event (about 300 kB per event) in order to analyse these data offline. In the first step we will try to extract from these data the relevant parameters for particle type and energy identification. The goal is to develop fast online data compression algorithms which will enable the extraction and histogramming of the relevant data online.

**5. Conclusion and Outlook**
In this article we have presented the basic design and operation parameters of the optical avalanche chamber (OPAC) - a particle track imaging device for dosimetry and dosimetry-related research. The device is able to measure the microscopic pattern of charge deposition in its gas medium and will provide a variety of physical data on which models of radiation action on biological entities are based.

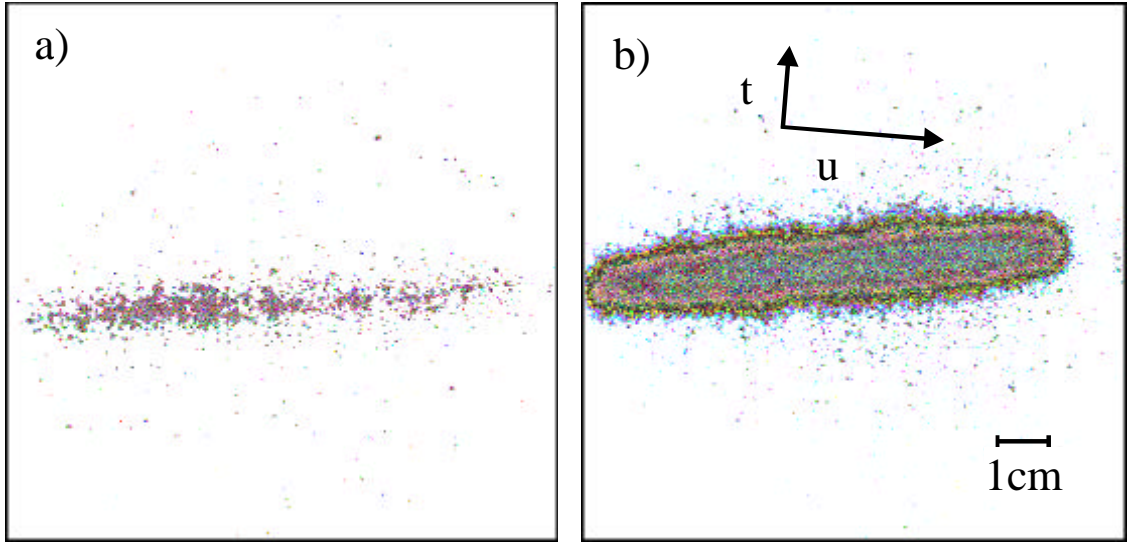

**Fig.11** *CCD image of **a**-particles in 10 hPa TEA with the OPAC obtained by*
*a) single-step charge and light amplification at an amplification voltage of $U_{A1} = 600V$*
*b) double-step charge and light amplification. Operation parameters: $1^{st}$ step $U_{A1} = 600\,V$ charge gain: $g_q = 5 \cdot 10^5$, $2^{nd}$ proportional scintillation stage: $U_{A2} = 800\,V$, light yield: $y_L = 2.5 \cdot 10^5$ photons).*

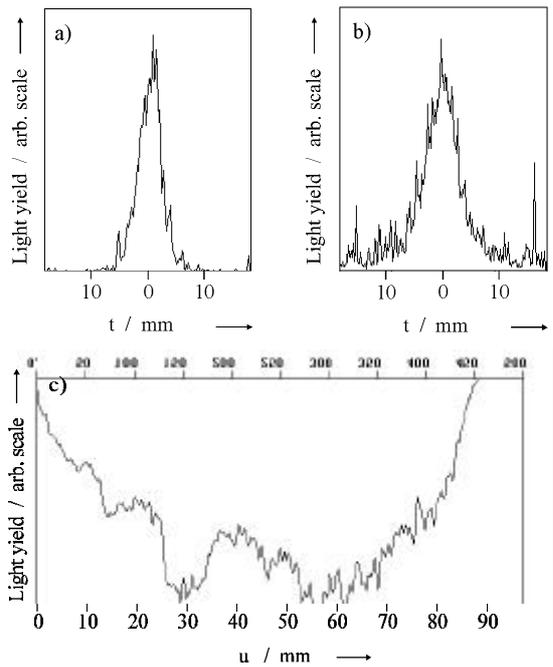

**Fig.12** *Ionization density distributions calculated from the images of fig.12. The coordinates u and t are defined in fig 12 b.*
*a) lateral ionization profile t with single-step amplification (correlated to fig. 12 a)*
*b) lateral ionization profile t with double-step amplification (correlated to fig. 12b)*
*c) longitudinal ionization profile u with double step amplification (correlated to fig.12b)*

The spatial resolution (pixel size) of about 3 mm by 3 mm in 10 hPa TEA corresponds to a pixel of 120 nm edge size at tissue density in a total sensitive volume of 4 $\mu m^3$. The position resolution can be improved at the expense of the size of the sensitive volume by reduction of the length of the drift path e.g. to 1 cm. At this drift path, the diffusion can be reduced by a factor of about three, and a lateral RMS resolution of less than 40 nm (scaled to tissue density) can be achieved. Such an operating mode is advantageous for microdosimetric measurements and investigations of ionization patterns relevant to radiation biology, where detection sensitivity is of minor importance.

In a further development stage, we foresee an improved detector design, where the chamber walls are made of tissue-equivalent material and the chamber gas composition is supplemented by organic additives to approximate tissue stoichiometry. For this modification the scintillation properties of these new gas mixtures have to be carefully investigated. With such a detector we could attempt to directly extract microdosimetric spectra or calculate proximity functions and deliver dose equivalent data in unknown radiation fields without field characterization.

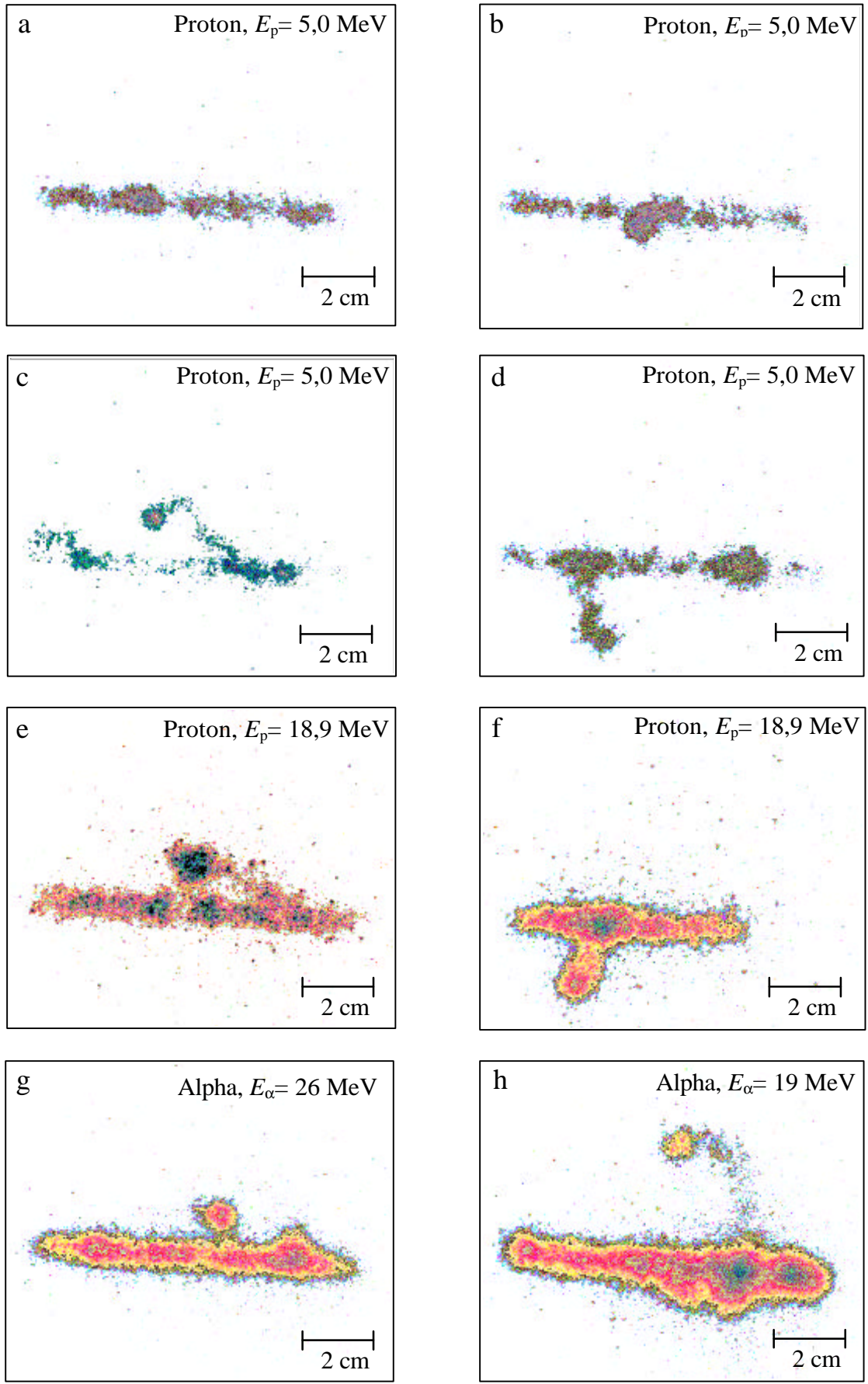

**Fig.13** *Images of proton and **a**-particle tracks in 10 hPa TEA. The images were obtained with particle injection parallel (a - e, g, h,) and 45° inclined (f) to the image projection plane. The flight direction of the particle is always from the right to the left side. The colour code between different pictures varies. Therefore no direct comparison of the ionisation density on the basis of equal colours is possible.*

The inelastic neutron reaction with $^3$He(n,p)t can be used to measure the spectral fluence of keV to several-MeV neutrons with proportional chambers. In a conventional spherical or cylindrical proportional chamber, the recoil particles deposit their energy in the chamber gas which contains several $10^5$ Pa of $^3$He. The signal measured is proportional to the sum of the nuclear reaction energy ($E_r$=770 keV) and the kinetic energy of the neutron. In cases in which the neutron capture occurs close to the wall or at high neutron energies, where the path length of the recoils exceeds the stopping length of the detector, the measurement underestimates the neutron energy (wall effect). Turner et al. proposed the application of a nuclear track imaging device [27]. A detector capable of measuring the track of both recoils, proton and triton, could provide the coordinates of the neutron-helium interaction locus and possible escapes of recoil ions from the sensitive volume (so that correction for the wall effect is possible). For high neutron energies, where the recoil cannot be stopped in the gas, one could measure the angle between the two recoils and calculate the neutron energy from the kinematic correlation between neutron, $^3$He and recoils.

For such an application the detector would need a $^3$He gas filling of at least $10^5$ Pa pressure to obtain sufficient sensitivity. Light emission of noble gas mixtures with TEA (Ar, Kr, Xe) was investigated by Suzuki [19], Sauvage [20] and Anderson [28]. It was shown that the emission spectra are determined by the TEA emission. The excitation energies and emission lines of the pure noble gases are in the vacuum UV range, thus TEA acts as a very efficient wavelength shifter. We therefore expect this behaviour to occur also in He / TEA. It was also shown [29] that the presence of He, with its very high first excitation levels (20.5 eV), does not alter significantly the field requirements for charge and light amplification. Therefore, the OPAC could be used without altering the UV optics or the parallel-grid amplification structure.

**Acknowledgements:**
We gratefully acknowledge the contribution of K. Tittelmeier to the development of the OPAC and the measurements. We thank D. Mugai and W.Wendt for their contributions to the experiment electronics and J. Asher, W. Heinemann and M. Klin for their technical support.

The work is supported by the European Commission (Contract FI4P-CT95-0024) and by the Foundation Mordoh Mijan de Salonique. A.Breskin is the W.P. Reuther Professor in the peaceful use of atomic energy.

**References:**

[1] Recommendations of the International Commission on Radiological Protection; ICRP Publication 60, Pergamon Press, Oxford

[2] Determination of Dose Equivalents Resulting from External Radiation Sources; International Commission on Radiation Units and Measurements, Bethesda, MD, ICRU 39

[3] H. G. Menzel, L. Lindborg, T. Schmitz, H. Schuhmacher, A. J. Waker; Radiat. Prot. Dosim. 29 (1989) 55

[4] S.R. Hunter; Nucl. Instrum. and Meth. A 260 (1987) 469

[5] J.E. Turner, R. Hunter, R.N. Hamm, H.A. Wright, G.S. Hurst and W.A. Gibson; Radiat. Prot. Dosim. 29 (1989) 9

[6] S.R. Hunter, W.A. Gibson, G.S. Hurst, J.E. Turner, R.N. Hamm and H.A. Wright; Radiat. Prot. Dosim. 52 (1994) 323

[7] T. Budd, M. Marshall and C.S. Kwok; Radiat. Res. 88 (1981) 228

[8] M. Marshall, G.P. Stonell and P.D. Holt; Radiat. Prot. Dosim. 13 (1985) 41

[9] G. Malamud, A. Breskin and R. Chechik; Nucl. Instrum. and Meth. A 307 (1991) 83

[10] A. Pansky, G. Malamud, A. Breskin and R. Chechik; Nucl. Instrum & Meth. A323 (1992) 294

[11] A. Breskin, R. Chechik, P. Colautti, V. Conte, A. Pansky, S. Shchemelinin, G. Talpo and G. Tornielli, Radat. Protec. Dosim. 61(1995)199.

[12] S. Shchemelinin, A. Breskin, R. Chechik, A. Pansky, P. Colautti; in „Microdosimetry", ed. by D.T. Goodhead, P. O'Neill, H.G. Menzel, ISBN 0-85404-737-9, (1997) 375

[13] U. Titt, A. Breskin, R. Chechik, V. Dangendorf, B. Großwendt and H. Schuhmacher; Radiat. Prot. Dosim. 70 (1996) 219

[14] G. Charpak, J.-P. Fabre, F. Sauli, M. Suzuki and W. Dominik; Nucl. Instrum. and Meth. A 258 (1987) 177 and references therein

[15] A. Breskin; Nucl. Phys. A 498 (1989) 457 and references therein

[16] D.F. Anderson, R. Bouclier, G. Charpak, S. Majewski and G. Kneller; Nucl. Instrum. and Meth. 217 (1983) 217

[17] P. Fonte, A. Breskin, G. Charpak, W. Dominik, F. Sauli; Nucl. Instrum. and Meth. A283 (1989) 658

[18] G. Charpak; Proc. Int. Symposium on Lepton and Photon Interactions at High Energies, Kyoto University (1985) 514

[19] M. Suzuki, P. Strock, F. Sauli, G. Charpak; Nucl. Intrum. and Meth. A 254 (1987) 556

[20] D. Sauvage, A. Breskin and R. Chechik; Nucl. Instrum. and Meth. A 275 (1989) 351

[21] V. Peskov, G .Charpak, W. Dominik, F. Sauli; Nucl. Instrum. and Meth. A 277 (1989) 347

[22] G. Charpak, W. Dominik, J.P. Fabre, J. Gaudean, V. Peskov, F. Sauli, M. Suzuki, A. Breskin, R. Chechik, D. Sauvage; IEEE Trans. Nucl. Sci. NS-35, (1988) 483

[23] A. Breskin, R. Chechik, D. Sauvage; Nucl. Instrum. and Meth. A 286 (1989) 251

[24] H.W. Funk; Proxitronic GmbH & Co. KG, Robert Bosch Straße 34, D-64625 Bensheim

[25] A. Breskin, G. Charpak, S. Majewski; Nucl. Instrum. and Meth. A220 (1984) 349

[26] A. Breskin, „Low Mass Detectors", in: Instrumentation for Heavy Ion Nuclear Research, Ed. D. Shapiro, Nuclear Science Research Conf. Series, Harwood Academic Publishers 7,75 (1985)

[27] J.E. Turner, R.N. Hamm, T.E. Houston, H.A. Wright, S.A. Gibson, G.S. Hurst; Radiat. Prot. Dosim. 32 (1990) 157

[28] D.F. Anderson, S. Kwan, V. Peskov and B. Hoeneisen; Nucl. Instrum. and Meth. A323 (1992) 626

[29] A. Breskin, R. Chechik, I. Frumkin; Nucl. Instrum. and Meth. A380 (1996) 562.